# Hybrid density functional study of band gap engineering of SrTiO$_3$ photocatalyst via doping for water splitting

Y. S. Hou[1,3], S. Ardo[2], and R. Q. Wu[3]*

[1] School of Physics, Sun Yat-Sen University, Guangzhou 510275, China

[2] Departments of Chemistry, Materials Science & Engineering, and Chemical & Biomolecular Engineering, University of California, Irvine, CA 92697-2025, USA

[3] Department of Physics and Astronomy, University of California, Irvine, CA 92697-4575, USA

## ABSTRACT

Perovskite SrTiO$_3$ (STO) is an attractive photocatalyst for solar water splitting, but suffers from a limited photoresponse in the ultraviolet spectral range due to its wide band gap. By means of hybrid density functional theory calculations, we systematically study engineering its band gap via doping 4$d$ and 5$d$ transition metals M (M=Zr, Nb, Mo, Tc, Ru, Rh, Pd, Hf, Ta, W, Re, Os, Ir and Pt) and chalcogen elements Y (Y=S and Se). We find that transition metal dopant M either has no effect on STO band gap or introduces detrimental mid-gap states, except for Pd and Pt that are able to reduce the STO band gap. In contrast, doping S and Se significantly reduces STO's direct band gap, thus leading to appreciable optical absorption transitions in the visible spectral range. Our findings provide that Pd, S and Se doped STO are potential promising photocatalysts for water splitting under visible light irradiation, thereby providing insightful theoretical guides for experiments to improve the photocatalytic activity of STO.

E-mail: wur@uci.edu

# I. INTRODUCTION

Photocatalytic water splitting provides an appealing way of producing hydrogen from water with solar energy [1-5]. Due to the match between the band alignment and the redox potential of water splitting reaction [6], the $ABO_3$-type perovskite $SrTiO_3$ (STO) is one of the most promising photocatalysts and has been widely studied experimentally and theoretically [5,7-21]. However, STO has a wide direct band gap of 3.75 eV [22], and hence its photoresponse is active only in the ultraviolet spectral range. It is crucial to engineer the band gap of STO photocatalysts to the visible light range, by doping STO with other elements [8,15,17,18,23,24], for which Rh and Sb, La are often used as dopants or co-dopants, respectively [7,11,16,19,24,25]. To guide the experimental efforts, several density functional theory (DFT) studies have been performed to investigate the electronic and optical properties of STO doped with different transition metals [16-19,26]. Nevertheless, conventional DFT calculations are known to underestimate the band gap of semiconductors and hence the dependability of existing predictions need to be examined by using higher level functionals such as the hybrid density functional which gives a more reliable band gap [27]. Furthermore, doping STO with anionic chalcogen elements S and Se, which are isovalent to oxygen but have stronger hybridization, has received much less attention so far. Given the condition that typical STO samples have appreciable oxygen vacancies [28-31], S or Se doping STO should be feasible. Compared with $4d$ and $5d$ transition metal dopants that usually have more $d$ electrons than Ti, S and Se dopants may broaden valence and conduction bands (hence shrinking the band gap) without introducing detrimental mid-gap states.

In this theoretical study, we employ hybrid density functional calculations to comprehensively investigate the electronic and optical properties of M-doped and Y-doped STO (M= Zr, Nb, Mo, Tc, Ru, Rh, Pd, Hf, Ta, W, Re, Os, Ir and Pt; Y=S and Se). For transition metal dopants, our calculations indicate that: 1) Zr and Hf have no obvious effect on the band gap; 2) Nb, Ta and W donate their electrons to the conduction bands of STO; 3) Pd and Pt reduce the band gap of STO; 4) Mo, Tc, Ru, Rh, Re, Os and Ir introduce mid-gap states. Interestingly, we find that doping a small amount of Se in STO, i.e., $SrTiSe_{0.0123}O_{2.9877}$, can already reduce the direct band gap of STO from 3.30 to 2.58 eV and hence noticeably enhance optical absorptions in the visible spectral range. This work provides a comprehensive theoretical understanding for rational band gap engineering of STO and, in particular, calling for attention of using Pd, S, and Se as dopants in STO for achieving high photocatalytic activity.

## II. COMPUTATIONAL DETAILS

All DFT calculations are carried out using the Vienna *Ab initio* Simulation Package (VASP) with the generalized gradient approximation formulated by Perdew, Burke and Ernzerhof (PBE) [32,33]. The projector-augmented wave pseudopotentials are used to describe the core-valence interaction [34,35]. More explicitly, we treat $Sr4s4p5s$, $Ti3d4s$ and $O2s2p$ as valance electrons; for dopants M (M= Zr, Nb, Mo, Tc, Ru, Rh, Pd, Hf, Ta, W, Re, Os, Ir and Pt) and Y (Y=S and Se), their valances electron are given at length in Table S1 in Supplementary Materials [36]. An energy cutoff of 350 eV is adopted for the basis expansion and atomic positions are fully relaxed until the force acting on each atom is smaller than 0.01 eV/Å. The Gaussian smearing is used and the energy convergence value is $10^{-6}$ eV in our calculations. To model the doped systems, we construct different supercells with a fixed lattice constant of 3.945 Å for the bulk STO. To correct the underestimated band gap in the conventional DFT calculations, we calculate the electronic and optical properties of the pristine and doped STO using the Heyd–Scuseria–Ernzerhof (HSE06) hybrid functional [27] with a mixing parameter $\alpha=0.25$ for the PBE and hybrid functionals. To reduce the heavy computational loads of the hybrid density functional calculations, band structures of pristine and doped STO are obtained by mean of the maximally localized Wannier function as implemented in the Wannier90 tool [37].

## III. RESULTS AND DISCUSSION

### 3.1 Geometry and electronic structure of pristine STO

Pristine STO has a centrosymmetric perovskite structure (Fig. 1a) with the Pm$\overline{3}$m symmetry group at room temperature. Ti atoms sit at the center of the octahedron that is formed by six $O^{2-}$ anions. Because of the octahedral crystal field generated by $O^{2-}$ anions, the five-fold degenerate 3$d$ orbitals of Ti atoms are split into low-lying three-fold degenerate $t_{2g}$ orbitals and high-lying two-fold degenerate $e_g$ orbitals. As $Ti^{4+}$ cations have no electrons in their 3$d$ orbitals, pristine STO is non-magnetic. However, 4$d$ and 5$d$ transition metal dopants may induce magnetization in STO, depending on the number of electrons filling their $d$ orbitals.

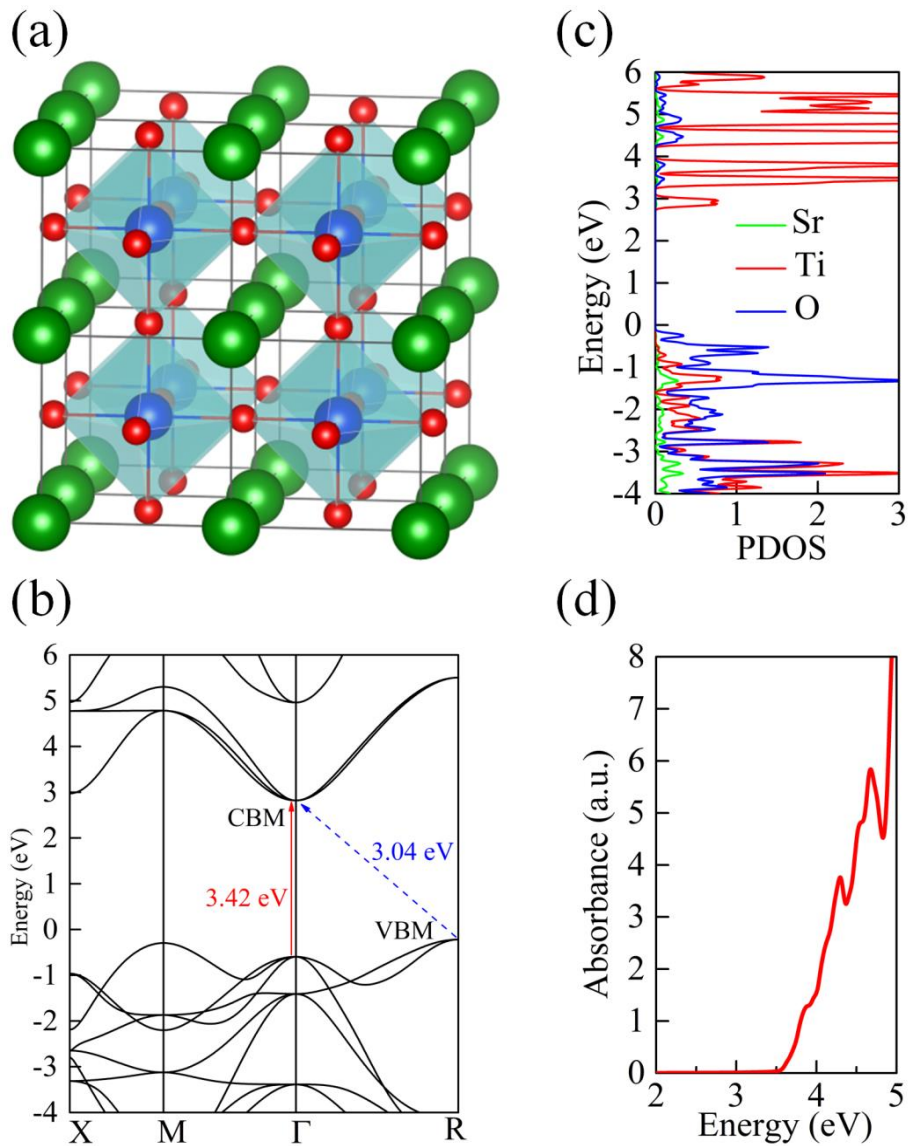

**Fig. 1.** *Crystal structure and electronic and optical properties of pristine STO. (a)*

*Crystal structure of the 2×2×2 supercell of STO. Ti, O, and Sr atoms are represented by blue, red and dark green balls, respectively. (b) Band structure, (c) PDOS (in units of state/eV) and (d) optical absorption of pristine STO. Fermi levels in (b) and (c) are set equal to zero.*

Fig. 1b shows the DFT calculated band structure of pristine STO. It is clear that the bulk of STO is a semiconductor with a wide indirect (direct) band gap of 3.04 (3.42) eV, consistent with the experimentally measured indirect (direct) band gap of 3.25 (3.75) eV [22]. Moreover, one can observe that the valence band maximum (VBM) is located at the R point while the conduction band minimum (CBM) is located at the Γ point. From the projected density of state (PDOS) as shown in Fig. 1c, one can see that: 1) the valence bands are mainly from $O^{2-}$ anions; 2) the conduction bands are mainly from $Ti^{4+}$ cations; 3) $Sr^{2+}$ cations mainly donate their electrons and their weights in the valence and conduction bands are small. As shown in Fig. 1d, the optical absorption of bulk STO starts at 3.50 eV, consistent with the experimentally observed optical absorption spectrum [38].

### 3.2 Electronic structures of STO doped with 4d and 5d transition metals M

Now we investigate the effects of $4d$ and $5d$ transition metal dopants: Zr, Nb, Mo, Tc, Ru, Rh, Pd, Hf, Ta, W, Re, Os, Ir and Pt. Considering that they have similar chemical properties and comparable ionic radii to Ti, we only study the case where one of $Ti^{4+}$ cations in the 2×2×2 supercell of STO is substituted by these atoms. For the convenience of discussions, we denote them as M@222-STO. Before proceeding to study the electronic properties of M@222-STO, let us pay attention to the band structure of the 2×2×2 supercell of STO (denoted as 222-STO). As shown in Fig. 2a, 222-STO appears to have a direct band gap of 3.04 eV due to the band folding. Depending on the number of electrons filling the $d$ orbitals of dopants, M@222-STO can be divided into two groups: one is nonmagnetic, and the other is magnetic.

We first examine the simplest nonmagnetic cases, Zr@222-STO and Hf@222-STO, in

which dopants Zr and Hf belong to the same group as Ti in the periodic table (Fig. 2k). Comparing the band structures of Zr@222-STO (Fig. 2b) and Hf@222-STO (Fig. S1 in Supplementary Materials [36]) to that of 222-STO (Fig. 2a), Zr and Hf basically have no effect on the band structure of the bulk STO. The PDOS curves in Fig. 2c and 2d indicate that the empty *d* orbitals of Zr and Hf lie higher in energy than those of Ti. Moreover, the filled *d* orbitals of Zr and Hf are delocalized and therefore more strongly hybridize with the 2*p* orbitals of oxygen.

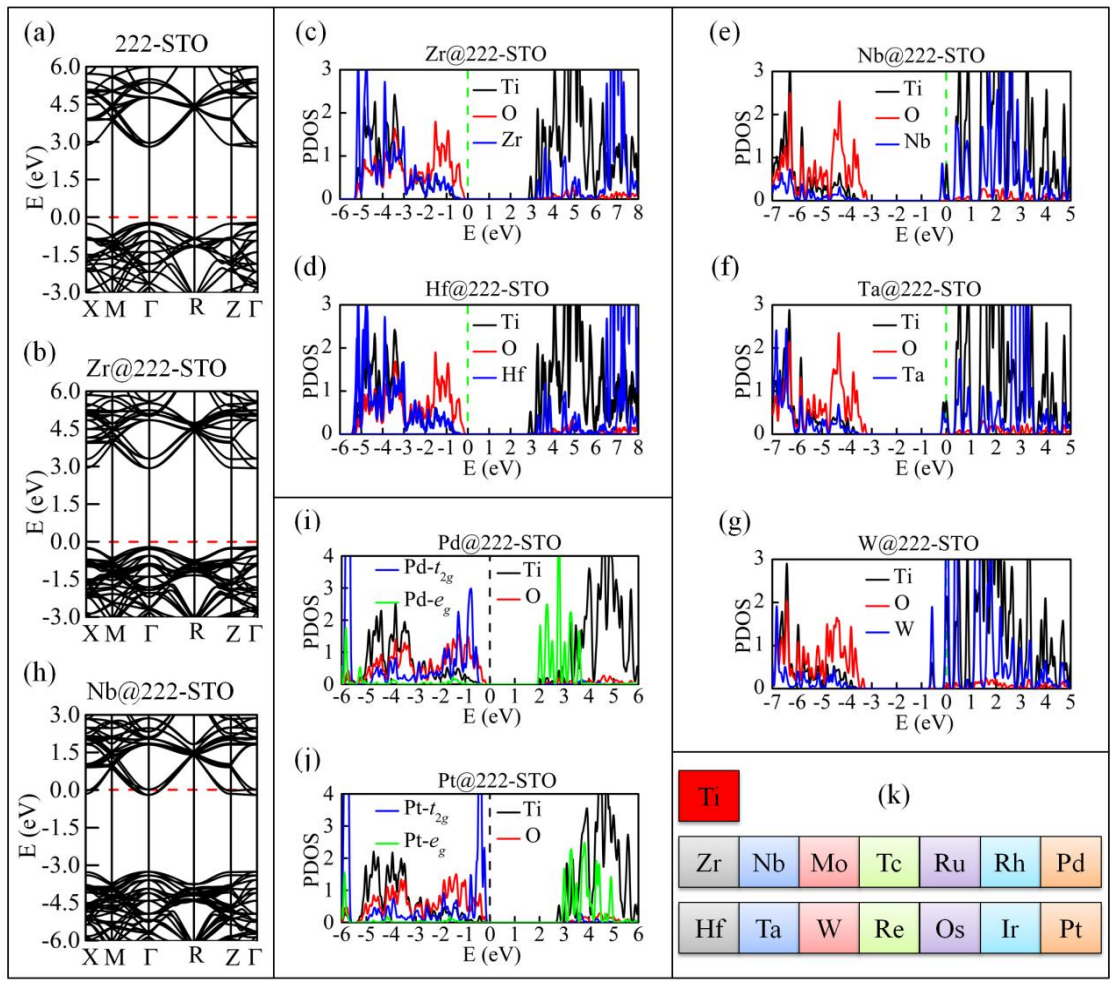

**Fig. 2**. *Band structures and PDOS of the nonmagnetic M@222-STO. (a), (b) and (h) are band structures of 222-STO, Zr@222-STO and Nb@222-STO, respectively. (c), (d), (e), (f), (g), (i) and (g) show PDOS (in units of state/eV) of Zr@222-STO, Hf@222-STO, Nb@222-STO, Ta@222-STO, W@222-STO, Pd@222-STO and Pt@222-STO, respectively. Fermi levels are set equal to zero in (a)-(j). (k) Relative position of the studied transition metal dopant M with respect to Ti in the periodic*

*table.*

Intuitively, dopant M should lose four electrons to oxygen neighbors and form the $M^{4+}$ state as a substitute to $Ti^{4+}$. However, our calculations show that Nb, Ta and W are exceptions, as they do not exhibit the expected magnetic moments of 1.0, 1.0 and 2.0 $\mu_B$, respectively, due to their unpaired electrons if they were in the +4 oxidation state. Conversely, our calculations show no magnetic moments in Nb@222-STO, Ta@222-STO and W@222-STO, even though nonzero magnetic moments are initially set in our calculations. Furthermore, their PDOS curves (Fig. 2e, 2f and 2g) and band structures (Fig. 2h and Fig. S2 [39]) clearly show that the $3d$ orbitals of Ti are partially filled. Evidently, this implies that the +4 oxidation state of Ti is reduced, and that the oxidation state of Nb, Ta and W is larger than +4. This more metallic nature of Nb@222-STO, Ta@222-STO and W@222-STO suggests that they might be poor at producing photovoltages.

Fig. 2i and 2j show the PDOS curves of the last two nonmagnetic cases: Pd@222-STO and Pt@222-STO. Our calculations show that the lower-lying $t_{2g}$ orbitals of $Pd^{4+}$ and $Pt^{4+}$ ions are occupied by six electrons whereas their higher-lying $e_g$ orbitals are empty. This unveils that the electronic configurations of both Pd and Pt dopants are $t_{2g}^6 e_g^0$ in the doped STO. Distinctly from the PDOSs of bulk STO, extra PDOS features appear in the conduction bands of Pd@222-STO (Fig. 2i) and in the valence bands of Pt@222-STO (Fig. 2j). Their band structures (Fig. 5a and Fig. S3 [39]) show that the direct band gaps of Pd@222-STO and Pt@222-STO are 2.24 and 3.00 eV, respectively. Hence, doping Pd and Pt, especially the former, can significantly reduce STO's direct band gap as desired.

Now let us investigate the magnetic M@222-STO where dopants M are Mo, Tc, Ru, Rh, Re, Os and Ir. Their PDOS curves (Fig. 3a-3g) manifest different majority and minority states, indicating nonzero magnetic moments in these systems. Our spin-polarized DFT calculations show that these systems have the following magnetic moments: 1.0 $\mu_B$ in Rh@222-STO and Ir@222-STO; 2.0 $\mu_B$ in Mo@222-STO, Ru@222-STO and Os@222-STO; 3.0 $\mu_B$ in Tc@222-STO and Re@222-STO. As discussed below, the M-dependent magnetic moments in these systems result from the interplay between the crystal field of oxygen octahedron and the delocalization nature of their 4$d$ and 5$d$ orbitals. Due to the large crystal field splitting between $t_{2g}$ and $e_g$ orbitals in the oxygen cage and the small Coulomb repulsion interaction of the delocalized 4$d$ and 5$d$ orbitals, electrons in the $d$ orbitals of M$^{4+}$ ions in magnetic M@222-STO prefer a low-spin state, thereby leading to the above-mentioned M-dependent magnetic moments (Fig. 3h).

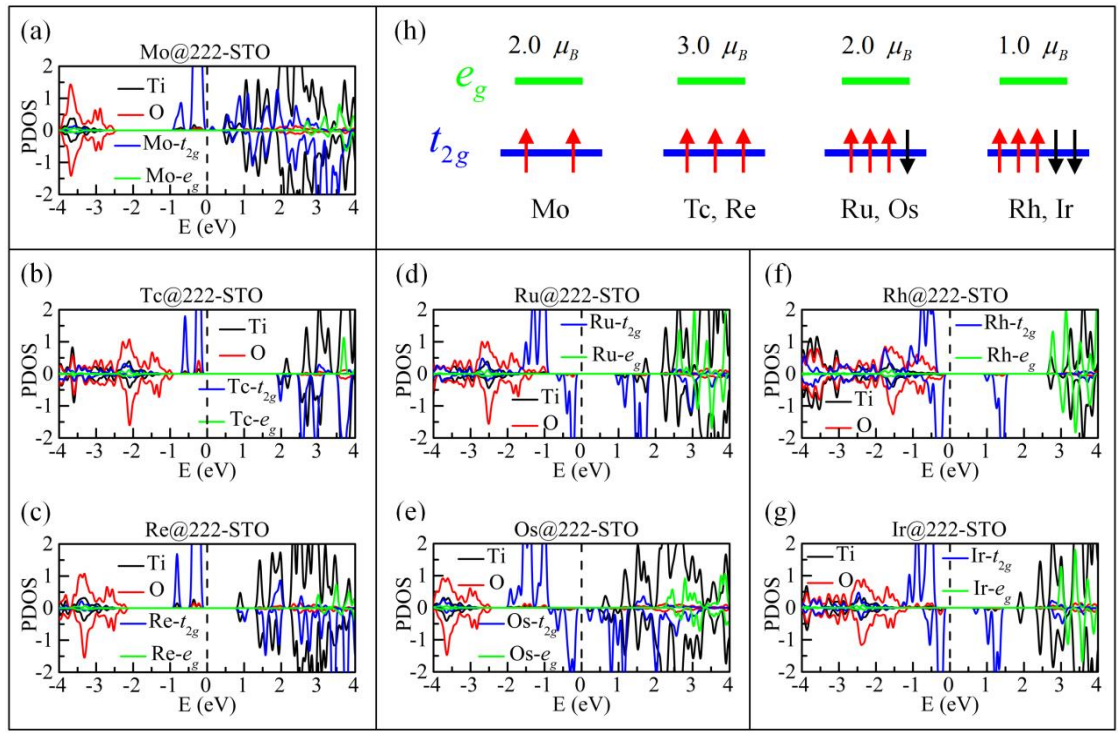

**Fig. 3.** *PDOS of M-doped STO with nonzero magnetic moments. (a)-(g) show the PDOS (in units of state/eV) of Mo@222-STO, Tc@222-STO, Re@222-STO, Ru@222-STO, Os@222-STO, Rh@222-STO, Ir@222-STO, respectively. Majority (minority) states have positive (negative) PDOS. (h) Electronic configurations in the d orbitals of magnetic dopant M$^{4+}$ (M=Mo, Tc, Ru, Rh, Re, Os and Ir) ions. Green and blue horizontal lines represent the low-lying three-fold degenerate $t_{2g}$ orbitals and*

*the high-lying two-fold degenerate $e_g$ orbitals, respectively. The red and black arrows denote the up-spin and down-spin electrons, respectively.*

Depending on dopants, the electronic properties of magnetic M@222-STO display rich variations as well. First of all, they all have mid-gap states as shown in Fig. 3a-3g. Some of these states are only from the majority spin channel (M=Mo, Tc and Re) (Fig. 3a-3c) whereas others are from both spin channels (M=Ru, Rh, Os and Ir) (Fig. 3d-3g). In particular, the majority spin $t_{2g}$ orbitals are fully occupied only in Tc@222-STO and Re@222-STO. Therefore, the valence bands contain contributions from Tc or Re, but the conduction bands are still the $3d$ orbitals of Ti in these two cases (Fig. 3b-3c). This is distinctly different from the other five systems studied with M=Mo, Ru, Rh, Os and Ir, in which both valence and conduction bands contain contributions from dopants.

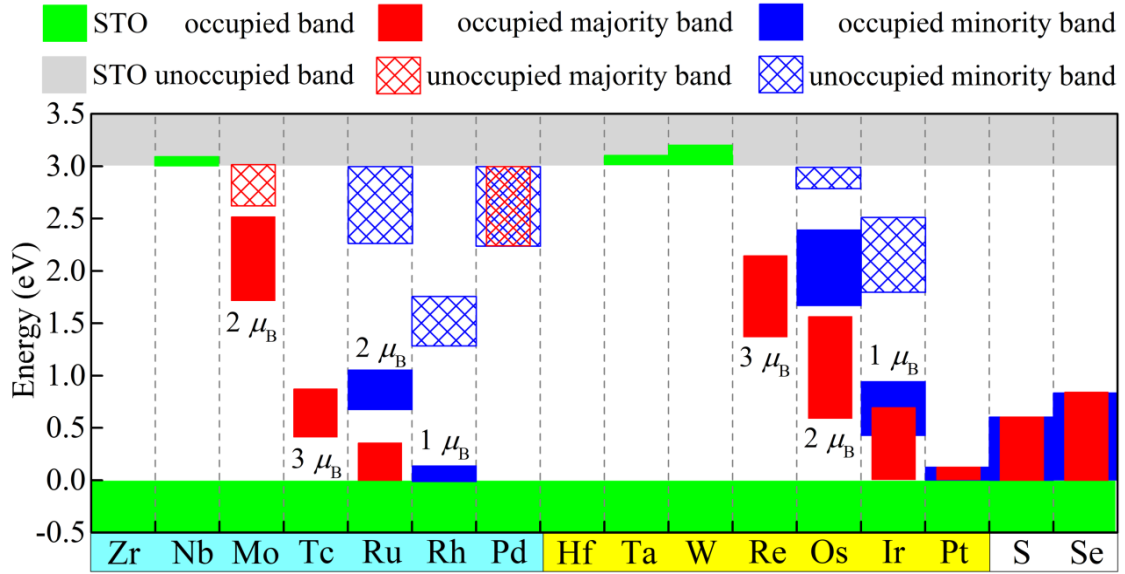

**Fig. 4.** *Summary of the bandwidths and band alignments of dopant M with respect to the valence and conduction bands of the STO host in M@222-STO. For completeness, S@222-STO and Se@222-STO are also included.*

From the above comprehensive information, we make a summary of the electronic properties of M@222-STO and show it in Fig. 4. Zr, Nb, Hf, Ta and W do not introduce mid-gap states, thereby having no noticeable effect on the band gap of the STO host. In contrast, Pd and Pt cause significant band gap reductions. Mo, Tc, Ru, Rh, Re, Os and Ir are complicated as they also produce magnetization. Taking Rh as an example, its occupied states lie in the valence bands of STO and its unoccupied states are separated from the conduction bands of STO in Rh@222-STO, consistent with previous theoretical studies [7,18]. Bands of Mo, Ru, Os and Ir doped STO have a similar layout as Rh@222-STO. Lastly, the occupied majority spin states of Re almost bisect the band gap of STO in Re@222-STO, but those of Tc are rather close to the top of valance bands of STO in Tc@222-STO.

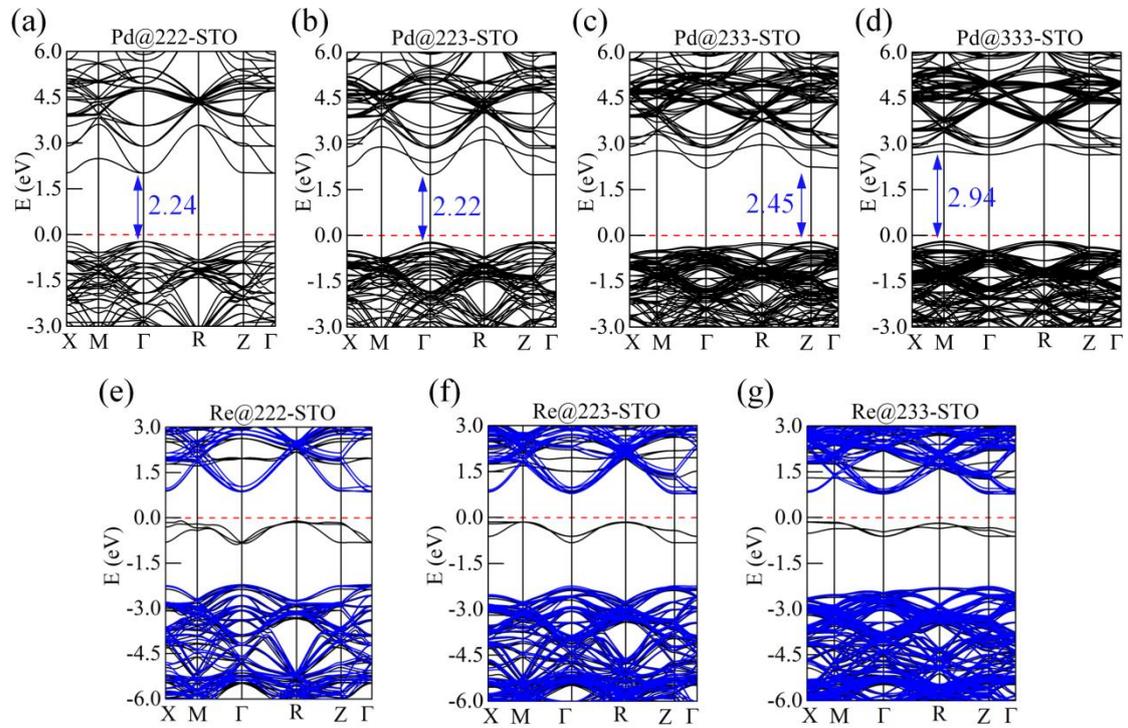

**Fig. 5.** *Effect of the concentration of dopant Pd/Re on the electronic properties of Pd/Re doped STO. (a)-(d) Band structures of Pd-doped STO and numbers give the direct band gap in units of eV. (e)-(g) Band structure of Re-doped STO. Majority and minority bands in (e)-(g) are plotted by the black and blue lines, respectively. Fermi levels are set equal to zero and indicated by the horizontal red dashed lines in (a)-(g).*

Now we examine how the concentration of dopants affects the electronic properties of STO. Taking Pd and Re as examples, we increase the supercell size of STO from 2×2×2 to 2×2×3, 2×3×3 and 3×3×3 and substitute only one Ti atom with Pd or Re in these supercells. As revealed by the band structures shown in Fig. 5a-5d, the band gap of Pd-doped STO increases gradually as the supercell size expands. This is understandable as the Pd-Pd interaction is reduced. Nevertheless, the direct band gap of Pd@333-STO is still smaller by 0.48 eV than that of the pristine STO within our calculations, even though the bandwidths of the occupied Pd states become very narrow. Similarly, the bandwidth of the occupied majority spin of Re states also shrinks in Re-doped STO as the supercell size increases (Fig. 5e-5g). Obviously, the concentration of dopants is an important factor of engineering the band gap of STO. We perceive that doping STO with high concentration Mo, Rh, Re and Ir may significantly improve its solar absorption efficiency as the midgap states may allow optical transitions from/to the valance and conduction bands of STO through dopant-host hybridizations.

### 3.3 Electronic structures of STO doped with chalcogen elements Y (Y=S and Se)

In this section, we study the electronic structures of STO doped with anionic dopants. Here we choose the chalcogen elements S and Se that are isovalent to oxygen. Fig. 6a and 6b show the PDOS curves of S@222-STO and Se@222-STO, respectively. We observe that states at the top of the valence bands of these two systems are mostly due to the S and Se dopants (Fig. 6c and 6d), since the $3p$ and $4p$ orbitals of dopants are much larger than the $2p$ orbitals of oxygen. This follows the similar strategy that we reported for O-alloying of pyrite $FeS_2$ to expand its band gap [40]. As a result, S@222-STO and Se@222-STO have significantly reduced band gaps: 2.40 and 2.16 eV, respectively. Fig. 6d-6g show the evolution of band structures of Se-doped STO as the concentration of Se decreases. Excitingly, we see that the band gap increases only by a small amount of 0.42 eV on going from Se@222-STO (2.16 eV) to Se@333-STO (2.58 eV). It is worthwhile to point out that the concentration of Se in Se@333-STO is already very low, i.e., 1.23%, comparable to that of Rh-doped STO in experiments [7].

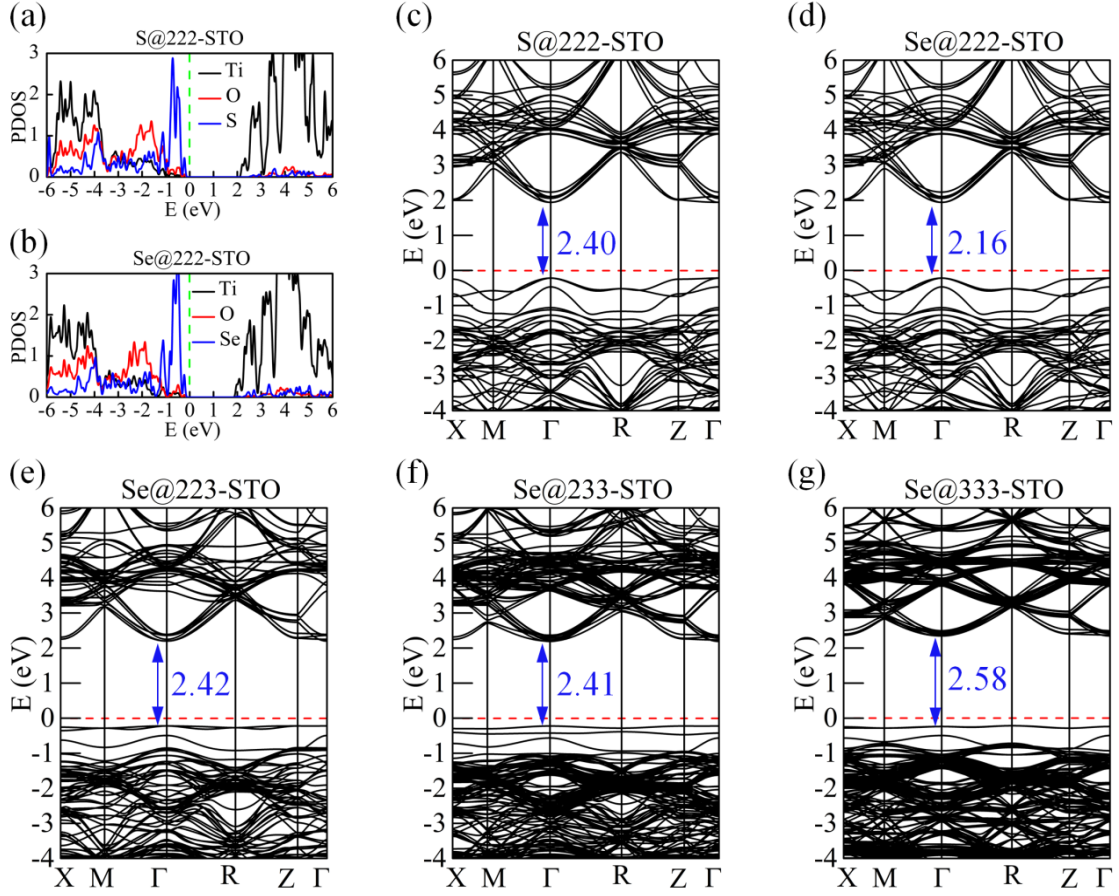

**Fig. 6.** *Electronic properties of STO doped with chalcogen elements Y (Y=S and Se). (a) and (c) show the PDOS (in units of state/eV) and band structure of S@222-STO. (b) and (d) same as (a) and (c) but for Se@222-STO. (e)-(g) show the band structures of Se-doped STO with different concentrations of Se. Numbers in (c)-(g) give the direct band gap in units of eV.*

### 3.4 Discussions

There are several key factors that we need to consider for tailoring the band gap of STO through doping. It is ideal to reduce the band gap of STO to below 2.0 eV, possibly via an increase in dopant concentration. Therefore, one must be concerned with the solubility and cost of dopants. For the strategy of expanding the widths of valence and conduction bands, we perceive that S, Se and Pd are good candidates. Previous studies [7,11,19] have shown that mid-gap states of $Rh^{4+}$ in Rh-doped STO are acceptor states and may trap charge carriers, thus being detrimental for photocatalysis. However, they may mediate optical absorption if these states become delocalized and have adequate wave function overlap with the conduction or valence states of STO host. Obviously, this could be a new strategy and only can be done with a high doping concentration. If the mid-gap bands are broad, they also give rise to a good mobility for photo-induced carriers, which mediates rapid separation of electrons and holes generated by light [5]. From the band widths and band alignments as summarized in Fig. 4, we see that many elements are promising candidates, and hope this motivates experimentalists to evaluate doped-STO containing non-traditional dopants.

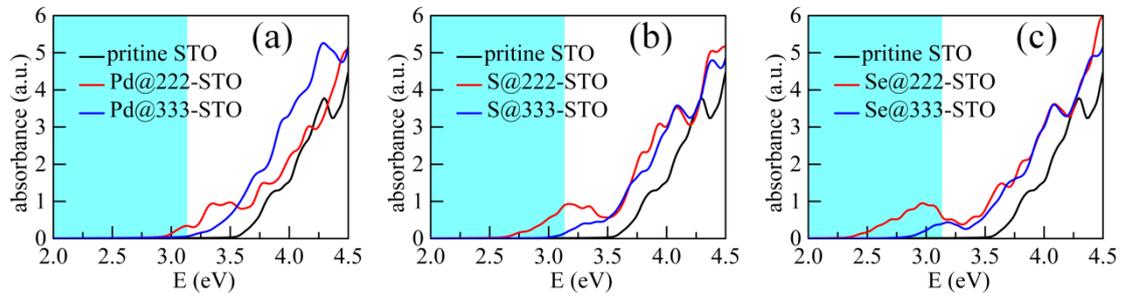

**Fig. 7.** *(a) Optical absorptions of Pd@222-STO and Pd@333-STO. (b) Optical absorptions of S@222-STO and S@333-STO. (c) Optical absorption of Se@222-STO and Se@333-STO. For comparisons, the optical absorption of the pristine STO is shown by the black lines in (a), (b) and (c). The cyan shades highlight the energy region that is in the range of the visible light.*

Here, we report the optical properties of the above-suggested photocatalysts, Pd, S, and Se doped STO. As seen in Fig. 7a, the optical absorptions of Pd@222-STO and Pd@333-STO start at photon energies of 2.95 and 3.13 eV. This shows that Pd-doped STO can be used for water splitting using near-ultraviolet light. Strikingly, the absorption thresholds of S and Se doped STO shift toward the visible light range, as shown in Fig. 7b and 7c, even with a small concentration of dopants (~1.3%). Therefore, Pd, S, and Se doped STO are indeed potential promising photocatalysts for water splitting under visible light irradiation.

Finally, we want to point out that there are various defects in experiments that may affect band gaps and optical properties of the doped STO as predicted in our work. It is shown that the common native defect, oxygen vacancy, can induce local moments in STO depending on the neighboring vacancy sites, vacancy-vacancy interaction, strain, and charge carrier density [41-43]. In this case, the oxygen-vacancy induced moments can couple to the moments of the magnetic $4d/5d$ dopants and, thus, the midgap states, band gaps may hence be modified to some extent. On the other hand, the dopants themselves may distribute non-uniformly in STO or even form small clusters. Obviously, these disorders also have effect on the band gaps and optical properties of the doped STO as discussed before [44]. Considering the complexity of these defect types and distributions, our present work cannot cover all aspects of the electronic properties of the doped STO, but rather provide useful insights for improving the photocatalytic performance of STO.

## IV. CONCLUSIONS

In summary, we study comprehensively how the band gap and optical properties of $SrTiO_3$ photocatalysts are engineered by doping $4d$ and $5d$ transition metals M (M=Zr, Nb, Mo, Tc, Ru, Rh, Pd, Hf, Ta, W, Re, Os, Ir and Pt) and chalcogen elements Y (Y=S and Se). Our study shows that transition metal dopants M have different effects on the band gap of STO depending on the electron numbers in their $d$ orbitals. First, isovalent Zr and Hf in place of Ti has no effect on the band gap of the STO host, and Nb, Ta and W donate electrons to the conduction bands of the STO host, making their electronic properties unsuitable for application in light-driven water splitting. Second, similar to Rh-doped STO, there are mid-gap states that are either occupied or unoccupied when doping Mo, Tc, Ru, Re, Os and Ir in STO. Based on previous studies of Rh-doped STO [7,18], codoping with other elements could be utilized to improve the photocatalytic activity in these systems. Lastly, it is rather exciting that the direct band gap of STO can be reduced by doping Pd, Pt, S and Se. Especially, Pd, S and Se doped STO have largely reduced direct band gaps and appreciable optical absorptions in the visible light range. Our work suggests that doping Pd, S and Se should be appealing to engineer STO as a photocatalyst for water splitting under visible light irradiation.

## ACKNOWLEDGMENTS

This work was supported by DOE-BES (Grant No. DEFG02-05ER46237) and UCI Research Seed Funding, and DFT calculations were performed on parallel computers at NERSC supercomputer centers. Yusheng Hou acknowledges the support from School of Physics, Sun Yat-Sen University.

(2014).

**Supplementary materials of "Hybrid density functional study of band gap engineering of SrTiO₃ photocatalyst via doping for water splitting"**

Y. S. Hou[1,3], S. Ardo[2], and R. Q. Wu[3]*

[1] School of Physics, Sun Yat-Sen University, Guangzhou 510275, China

[2] Departments of Chemistry, Materials Science & Engineering, and Chemical & Biomolecular Engineering, University of California, Irvine, CA 92697-2025, USA

[3] Department of Physics and Astronomy, University of California, Irvine, CA 92697-4575, USA

Email: wur@uci.edu

**S1. Tabel I. Valence electrons of dopant M and Y. Here VEs is short for valance electrons.**

| Dopant M | VEs | Dopant M | VEs | Dopant Y | VEs |
|---|---|---|---|---|---|
| Zr | $4d5s$ | Hf | $5d6s$ | S | $3s3p$ |
| Nb | $4s4p4d5s$ | Ta | $5d6s$ | Se | $4s4p$ |
| Mo | $4d5s$ | W | $5d6s$ | | |
| Tc | $4d5s$ | Re | $5d6s$ | | |
| Ru | $4d5s$ | Os | $5d6s$ | | |
| Rh | $4d5s$ | Ir | $5d6s$ | | |
| Pd | $4d5s$ | Pt | $5d6s$ | | |

**S2. Band structure of Hf@222-STO**

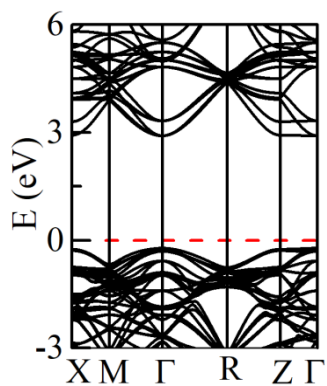

**Figure S1**. DFT calculated band structure of Hf@222-STO. Fermi level is set to be zero and indicated by horizontal red line.

**S3. Band structures of Ta@222-STO and W@222-STO**

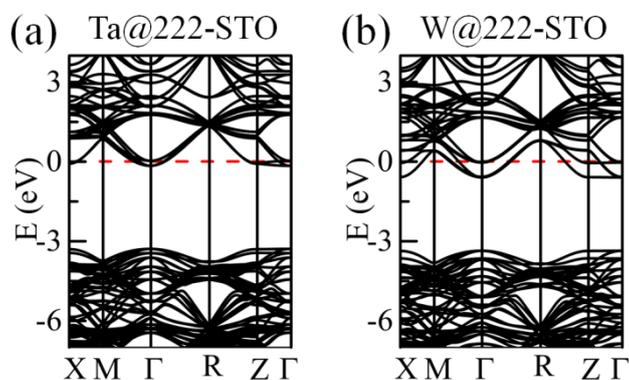

**Figure S2**. DFT calculated band structure of (a) Ta@222-STO and (b) W@222-STO. Fermi levels are set to be zero and indicated by horizontal red line in (a) and (b).

## S4. Band structure of Pt@222-STO

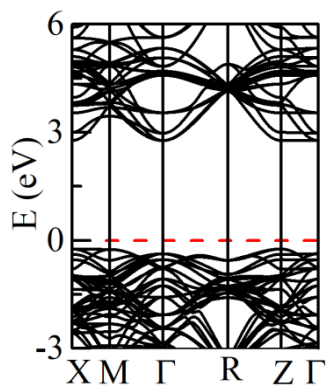

**Figure S3**. DFT calculated band structure of Pt@222-STO. Fermi level is set to be zero and indicated by horizontal red line.

## S5. Band structures of M@222-STO (M=Mo, Tc, Ru, Rh, Os, and Ir)

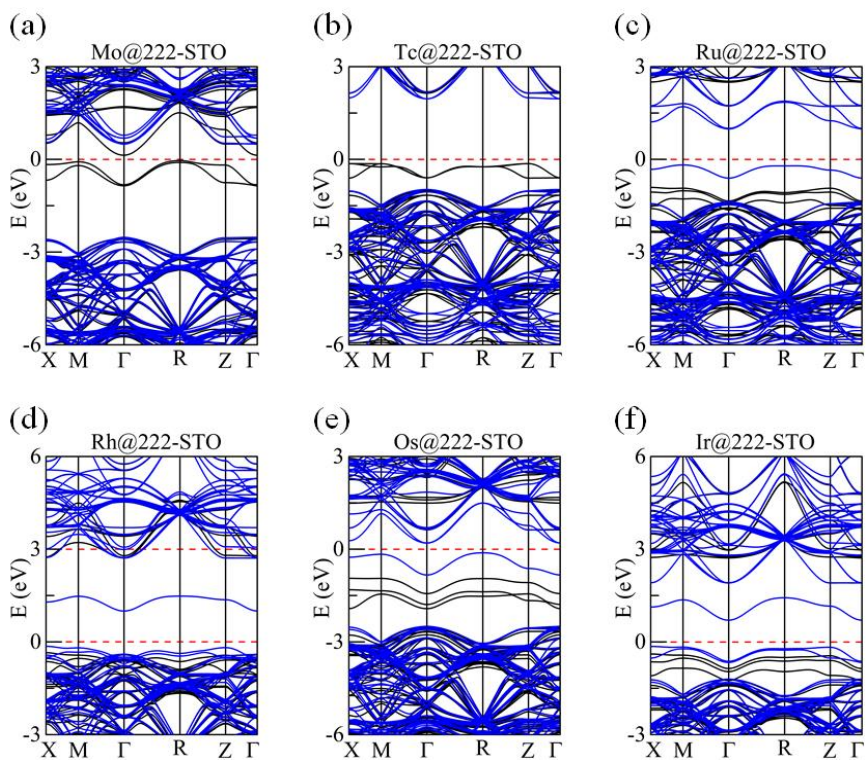

**Figure S4**. DFT calculated band structure of M@222-STO (M=Mo, Tc, Ru, Rh, Os, and Ir). Fermi level is set to be zero and indicated by horizontal red line. The majority and minority bands are shown by the black and blue lines.